\documentclass[preprint,prd,eqsecnum,amsfonts,floatfix,nofootinbib]{revtex4}
\usepackage[pdftex]{graphicx}
\usepackage{hyperref}
\usepackage{amsmath}
\usepackage{aeitensor}
\usepackage{multirow}
\usepackage{url}
\usepackage{color}
\newcommand{\sref}[1]{section~\ref{#1}}
\newcommand{\fref}[1]{figure~\ref{#1}}
\newcommand{\tref}[1]{table~\ref{#1}}

\newcommand{\abs}[1]{\left\lvert#1\right\rvert}
\newcommand{\un}[1]{\text{\,#1}}
\newcommand{\mc}[1]{\mathcal{#1}}
\newcommand{\mf}[1]{\mathfrak{#1}}
\newcommand{\wt}[1]{\widetilde{#1}}
\newcommand{\F}{\mathcal{F}}

\DeclareMathOperator{\sinc}{sinc}
\DeclareMathOperator{\erfc}{erfc}

\newcommand{\khat}{\zhat}
\newcommand{\nhat}{\widehat{n}}

\newcommand{\zhat}{\widehat{k}}

\newcommand{\mbf}[1]{\ensuremath{\mathchoice{\mbox{\boldmath$\displaystyle#1$}}
{\mbox{\boldmath$\textstyle#1$}}
{\mbox{\boldmath$\scriptstyle#1$}}
{\mbox{\boldmath$\scriptscriptstyle#1$}}}}

\newcommand{\tens}[1]{\aeitensor{#1}}

\newcommand{\expected}[1]{E\left[ #1 \right]}

\newcommand{\synthLISA}{\text{synthLISA}}
\newcommand{\LISAsim}{\text{LISAsim}}

\newcommand{\SFT}{\text{SFT}}
\newcommand{\LWL}{\text{LWL}}

\newcommand{\doppler}{\theta}
\newcommand{\ddoppler}{\Delta\doppler}
\newcommand{\Amp}{\A}
\newcommand{\dAmp}{\Delta\Amp}
\newcommand{\Fisher}{\bar{\Gamma}}
\newcommand{\lon}{\lambda}
\newcommand{\lat}{\beta}

\newcommand{\sig}{{\mathrm{s}}}

\newcommand{\ampErr}{\epsilon_\Amp}
\newcommand{\dopErr}{\epsilon_\doppler}
\newcommand{\delA}{{\delta_\Amp}}
\newcommand{\phiA}{{\phi_\Amp}}

\newcommand{\cand}{{\mathrm{c}}}
\newcommand{\Akey}{\A_{\sig}}
\newcommand{\Acand}{\A_{\cand}}

\newcommand{\A}{\mc{A}}
\newcommand{\M}{\mc{M}}

\newcommand{\dcc}{LIGO-P080087-v5}
\newcommand{\Tlight}{T}

\newcommand{\coord}{coordinate}

\newcommand{\coinc}{coincidence}

\def\commitDATE{ Thu Jan 7 17:04:05 2010 -0500}

\newcommand{\numFound}{1989}
\newcommand{\numFoundLWL}{1704}
\newcommand{\numFalse}{5}

\newcommand{\twoFMin}{69.4}
\newcommand{\numBright}{59401}
\newcommand{\numForty}{6586}

\newcommand{\RAvRAAvsigAmu}{1.56}
\newcommand{\RAvRAAvmuF}{0.01}
\newcommand{\RAvRAAvsigF}{1.68}

\newcommand{\RAvRAAvnumSigs}{1989}
\newcommand{\RAvRAAvnumAmuSigs}{7956}

\newcommand{\RAvLWLvsigAmu}{2.21}

\newcommand{\LWLvLWLvsigAmu}{15.22}

\newcommand{\subtNumFound}{3419}
\newcommand{\subtNumFalse}{29}
\newcommand{\subtRounds}{seven}

\newcommand{\numMissedFortyMTJPL}{49}
\newcommand{\numFoundMTJPL}{18084}

\begin{document}
\title[Galactic WD Binary Search in LISA w/$\F$-stat template bank]
{Searching for Galactic White Dwarf Binaries in Mock LISA
  Data using an $\F$-Statistic Template Bank}
\author{John T Whelan}
\email{john.whelan@astro.rit.edu}
\affiliation{Center for Computational Relativity and Gravitation
  and School of Mathematical Sciences, Rochester Institute of Technology,
  85 Lomb Memorial Drive, Rochester, NY 14623, USA}
\author{Reinhard Prix}
\email{reinhard.prix@aei.mpg.de}
\affiliation{Max-Planck-Institut f\"{u}r Gravitationsphysik
  (Albert-Einstein-Institut), D-30167 Hannover, Germany}
\author{Deepak Khurana}
\affiliation{Indian Institute of Technology, Kharagpur, West Bengal
  721302, India}
\date{\commitDATE}
\begin{abstract}
  We describe an $\F$-statistic search for continuous gravitational waves from
  galactic white-dwarf binaries in simulated LISA Data.
  Our search method employs a hierarchical template-grid based exploration
  of the parameter space.  In the first stage, candidate sources are
  identified in searches using different simulated laser signal
  combinations (known as TDI variables).
  Since each source generates a primary maximum near its true
  ``Doppler parameters'' (intrinsic frequency and sky position) as
  well as numerous secondary maxima of the $\F$-statistic in Doppler
  parameter space, a search for multiple sources needs to distinguish
  between true signals and secondary maxima associated with other,
  ``louder'' signals.  Our method does this by applying a {\coinc}
  test to reject candidates which are not found at nearby parameter
  space positions in searches using each of the three TDI variables.
  For signals surviving the {\coinc} test, we
  perform a fully coherent search over a refined parameter
  grid to provide an accurate parameter
  estimation for the final candidates.  Suitably tuned, the pipeline is
  able to extract {\numFound} true signals with only {\numFalse} false
  alarms.  The use of the rigid adiabatic approximation allows
  recovery of signal parameters with errors comparable to statistical
  expectations, although there is still some systematic excess with
  respect to statistical errors expected from Gaussian noise.  An experimental
  iterative pipeline with {\subtRounds} rounds of signal subtraction and
  re-analysis of the residuals allows us
  to increase the number of signals recovered
  to a total of {\subtNumFound} with {\subtNumFalse} false alarms.
\end{abstract}
\preprint{\dcc}
\maketitle

\section{Introduction}
\label{s:intro}

The Mock LISA Data Challenges (MLDCs)~\cite{mldc:_homepage} have the
purpose of encouraging the development of LISA data-analysis tools and
assessing the technical readiness of the community to perform
gravitational-wave (GW) astronomy with LISA.  The rounds completed so
far have been labelled MLDC1~\cite{Arnaud:2007vr},
MLDC2~\cite{Babak:2007zd}, MLDC1B~\cite{Babak:2008sn}, and
MLDC3~\cite{Babak:2008sn,Babak:2009cj}.  The challenges have
consisted of several data-sets containing different types of simulated
sources and LISA noise, including quasi-periodic signals from white-dwarf
binaries (WDBs).  In this paper we describe an analysis performed on
MLDC2 data, using an improved version of the pipeline that we originally
applied in our MLDC2 entry\cite{Babak:2007zd,MLDC2Poster}

GW signals from WDBs will be long-lasting and
\mbox{(quasi-)}monochromatic with slowly-varying intrinsic frequency
$f(\tau)$; in this sense they belong to the class of \emph{continuous
  GWs}.
In the case of ground-based detectors the typical sources of
continuous GWs are spinning neutron stars with non-axisymmetric
deformations.  One of the standard tools developed for these searches
is the $\F$-statistic.
We have applied this method in our MLDC searches, adapting the
LAL/LALApps~\cite{lalapps} search code \texttt{ComputeFStatistic\_v2}
used within the LIGO Scientific Collaboration to search for periodic
GW signals in data from ground-based detectors such as LIGO and
GEO\,600, e.g.\ see \cite{2008CQGra..25w5011W}.

MLDC1 and MLDC1B contained data sets with a relatively small number of
simulated WDB signals, and the results of our searches on those data
are reported elsewhere~\cite{Prix:2007zh,Whelan:2008zz}.  The MLDC2 data-set
contains a full simulated galaxy of WDB signals, with the
challenge being to extract as many of these signals as possible.  One
approach, used by Crowder et al~\cite{Crowder:2006eu,Crowder:2007ft},
is to fit the
overall signal with a multi-source template.  Our analysis instead
applies the traditional method of searching for individual sources.
An important challenge in that regard is to distinguish secondary
maxima in parameter space from primary peaks of true signals.  We
accomplish this through a hierarchical pipeline, which follows up
candidates found in {\coinc} between searches carried out with
different LISA observables.

The plan for the rest of this paper is as follows: In \sref{s:method}
we review the fundamentals of the $\F$-statistic search as applied to
WDB signals in mock LISA data.  In \sref{s:pipeline} we describe our
pipeline including the {\coinc} condition used to distinguish true
signals from secondary maxima, and the estimation of expected
statistical errors in the signal parameters.  In \sref{s:eval} we
describe some of the techniques used to evaluate the effectiveness of
our pipeline: the post-hoc classification of candidates into found
signals and false alarms, and the discrepancies between the candidate
parameters returned by our pipeline and the simulated values.  In
\sref{s:results} we describe the results of our pipeline in its
optimal configuration and compare those with the results obtained
using less sophisticated models of the LISA response.  In
\sref{s:subtract} we present the results of an iterative program in
which the signals found by the pipeline are subtracted from the data
stream and then the pipeline is re-run on the residuals.

\section[WDB search method]
{Search method for continuous signals from white-dwarf binaries}

\label{s:method}

\subsection{The $\F$-statistic}

\label{ss:method-fstat}

The $\F$-statistic was originally developed
in~\cite{Jaranowski:1998qm}, extended to the multi-detector case
in~\cite{Cutler:2005hc}, and generalized to the full Time-Domain
Interferometry (TDI)~\cite{TDI:_1999} framework for
LISA in~\cite{Krolak:2004xp}.  The formalism for our application of this
method to mock LISA data has been described in \cite{Prix:2007zh} and
\cite{Whelan:2008zz}, to which the reader is referred for details.
Here we review the fundamentals of the method relevant to the current
application.

The signal received from a monochromatic GW source like a white-dwarf
binary with negligible orbital evolution can be characterized by seven
parameters.  The three \emph{Doppler parameters} are the intrinsic
frequency $f$ and two {\coord}s describing the sky location, such as
galactic latitude $\lat$ and longitude $\lon$, and can be denoted as
$\doppler \equiv \{f, \lat, \lon\}$.  The four \emph{amplitude
  parameters} are the overall GW amplitude $h_0$, the inclination
angle $\iota$ of the orbital plane, the polarization angle $\psi$, and
the initial phase $\phi_0$.  One set of convenient combinations
$\Amp^\mu = \Amp^\mu(h_0, \iota, \psi, \phi_0)$ is
\begin{subequations}
  \begin{align}
    \Amp^1 &= A_{+}\cos\phi_0\cos 2\psi - A_{\times}\sin\phi_0\sin 2\psi\,,
    \\
    \Amp^2 &= A_{+}\cos\phi_0\sin 2\psi + A_{\times}\sin\phi_0\cos 2\psi\,,
    \\
    \Amp^3 &= - A_{+}\sin\phi_0\cos 2\psi - A_{\times}\cos\phi_0\sin 2\psi\,,
    \\
    \Amp^4 &= - A_{+}\sin\phi_0\sin 2\psi + A_{\times}\cos\phi_0\cos 2\psi\,,
  \end{align}
\end{subequations}
where $A_{+}=h_0(1+\cos^2\iota)/2$ and $A_{\times}=h_0\cos\iota$.
Using these combinations, it is possible to write the signal received
in a detector $I$ with instrumental noise $n^{I}(t)$ as
\begin{equation}
  x^{I}(t) = n^{I}(t) + \Amp^\mu\, h^{I}_\mu(t; \doppler)\,,
\end{equation}
where we introduce the convention of an implicit sum $\sum_{\mu=1}^4$
over repeated indices $\mu, \nu$, and the form of the template
waveforms $h^{I}_\mu(t; \doppler)$ depends on the Doppler parameters
and the specifics of the detector, such as orientation and motion as a
function of time.

Following the notation of~\cite{Cutler:2005hc,Prix:2006wm}, we
write the different data-streams $x^I(t)$ as a vector $\mbf{x}(t)$,
and we define the standard multi-detector (with uncorrelated noise)
scalar product  as
\begin{equation}
  \label{e:innprod}
  (\mbf{x}|\mbf{y}) = \sum_{\alpha} \sum_{I} \int_{-\infty}^{\infty}
  \wt{x}_{\alpha}^{I*}(f)\, [S_{\alpha\,I}(f)]^{-1}
  \, \wt{y}_{\alpha}^I(f)\, df \,.
\end{equation}
Here we have broken up the observation time into intervals labelled by
$\alpha$, $\wt{x}_{\alpha}$ is the Fourier-transform of the data in
the $\alpha$th time interval, $x^*$ denotes complex conjugation, and
$\{S_{\alpha\,I}(f)\}$ is the two-sided noise power spectral
density appropriate to the $\alpha$th time interval.
We search for a signal $\{\Amp_\sig, \doppler_\sig\}$ by seeking the parameters
$\{\Amp_\cand, \doppler_\cand\}$
which maximize the log-likelihood ratio
\begin{equation}
  \hspace*{-1cm}
  L(\mbf{x}; \Amp, \doppler )
  = (\mbf{x}|\mbf{h}) - \frac{1}{2}(\mbf{h}|\mbf{h})
  = \mc{A}^\mu (\mbf{x}|\mbf{h}_\mu)
  - \frac{1}{2}\mc{A}^\mu (\mbf{h}_\mu|\mbf{h}_\nu) \mc{A}^\nu\,.
\end{equation}
Defining
\begin{equation}
  x_\mu(\doppler) \equiv (\mbf{x}|\mbf{h}_\mu)\,,\quad\mbox{and}\quad
  \mc{M}_{\mu\nu}(\doppler) \equiv (\mbf{h}_\mu|\mbf{h}_\nu)\,,
  \label{e:Ametric}
\end{equation}
we see that $L$ is maximized for given $\doppler$ by the amplitude
estimator $\Acand^\mu = \mc{M}^{\mu\nu}x_\nu$, where
$\mc{M}^{\mu\nu}$
is the inverse matrix of $\mc{M}_{\mu\nu}$. Thus the detection
statistic $L$, maximized over the amplitude parameters $\Amp$ is
\begin{equation}
  2\F(\mbf{x}; \doppler) \equiv \, x_\mu \, \mc{M}^{\mu\nu} \,x_\nu\,,
  \label{e:Fstat}
\end{equation}
which defines the (multi-detector) $\F$-statistic. One can show that
the expectation in the perfect-match case $\doppler = \doppler_\sig$
is $E[2\F(\doppler_\sig)] = 4 + \abs{\mc{A}_\sig}^2$, where we used the
definition
\begin{equation}
  \abs{\mc{A}}^2 \equiv \mc{A}^\mu \mc{M}_{\mu\nu}(\doppler_\sig) \mc{A}^\nu\,,
\end{equation}
for the norm of a 4-vector $\mc{A}^\mu$, using $\mc{M}_{\mu\nu}$ as a
\emph{metric} on the amplitude-parameter space.
Note that $\abs{\mc{A}_\sig}$ is the (optimal) signal-to-noise ratio
(SNR)  of the true signal $\{\mc{A}_\sig, \doppler_\sig\}$.

\subsection{Modelling the LISA response}

\label{ss:method-resp}

The MLDC data were generated by two different programs: Synthetic
LISA~\cite{synthLISA} simulates a detector output consisting of
Doppler shifts of the LISA lasers due to relative motion of the
spacecraft, while LISA Simulator~\cite{LISAsim} simulates the phase
differences between laser light following different paths between the
spacecraft.\footnote{Our pipeline was constructed to handle either
  LISA simulator or synthetic LISA data, but for concreteness the
  results in this paper were all generated from the synthetic LISA
  data.}
In both cases the underlying variables are combined with
appropriate time shifts to form TDI observables which cancel the
(otherwise dominating) laser frequency
noise~\cite{TDI:_1999,Tinto:2003vj,Krolak:2004xp}.  One choice of such
TDI quantities is the set of three observables $\{X, Y, Z\}$.  These
observables, which can be thought of as representing the output of
three virtual ``detectors'' (which we label with the index $I$), are
related to the gravitational wave tensor $\tens{h}$ through the
detector ``response'', which can be modelled at different levels of
accuracy.  Our current approach uses the \emph{rigid adiabatic
approximation}~\cite{Rubbo:2003ap}, but we also consider the
\emph{long-wavelength limit} (LWL).
In the LWL approximation the reduced wavelength
$c/(2\pi f)$ is assumed to be large compared to the distance $L$
between the spacecraft, which corresponds to a light-travel time of
$\Tlight = L/c \sim 17\un{s}$ (assuming equal arm-lengths), and so this
approximation requires $f \ll 10\un{mHz}$.
These alternatives and their consequences are considered in more
detail in \cite{Whelan:2008zz}, but here we summarize the relevant
approximations as they apply to our search.

\begin{figure}
  \centering
  \includegraphics[width=0.5\textwidth]{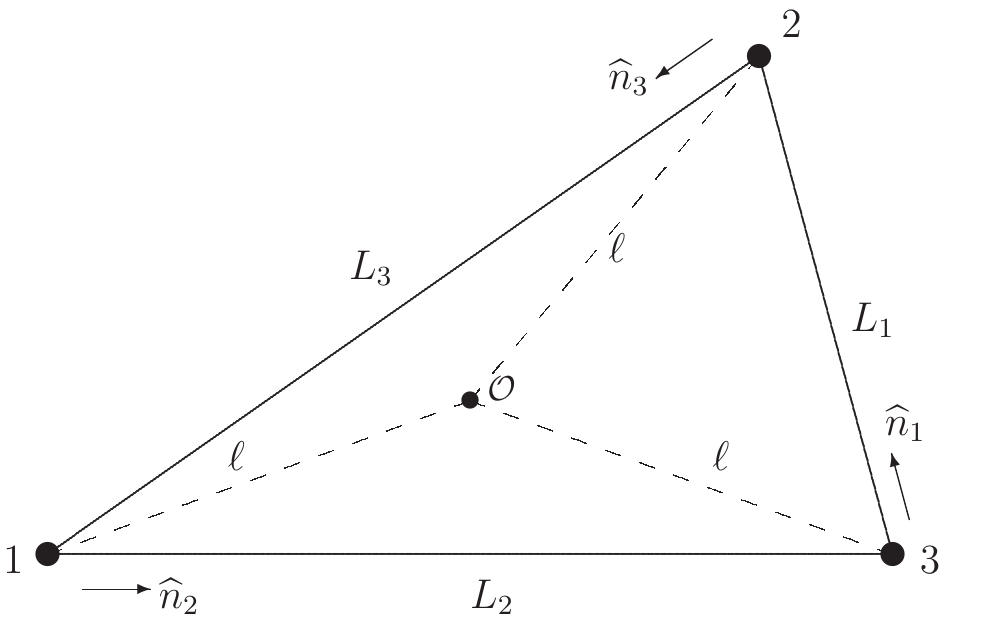}
  \caption{LISA configuration and TDI conventions used.}
  \label{f:LISA}
\end{figure}

It is convenient to describe the ``response'' of a gravitational wave
detector in the frequency domain in terms of a response function
$R(f)$, relating the detector output to a ``strain'' more closely
connected to the metric perturbation tensor $\tens{h}$, so that
\begin{equation}
  \wt{X}(f) = \frac{\wt{h}^{X}(f)}{R(f)}
  = \frac{\tens{d}^{X} : \wt{\tens{h}}(f)}{R(f)}
  \,,
\end{equation}
where $:$ denotes the contraction of both tensor indices.
In the long-wavelength limit,
\begin{subequations}
  \begin{align}
    R^{\synthLISA}(f) \approx R^{\synthLISA}_{\LWL}(f)
    &= \left(\frac{1}{4\pi f \Tlight}\right)^2\,,
    \\
    R^{\LISAsim}(f) \approx R^{\LISAsim}_{\LWL}(f)
    &= i \frac{1}{4\pi f \Tlight} \,,
  \end{align}
\end{subequations}
and $\tens{d}^{X}\approx\tens{d}^{X}_{\LWL}\equiv(\nhat_2\otimes\nhat_2 -
\nhat_3\otimes\nhat_3)/2$ is the usual LWL response tensor for a GW
interferometer with arms $\nhat_2$ and $\nhat_3$. The analogous
expressions for $Y$ and $Z$ are obtained by cyclic permutations of the
indices $1\rightarrow 2\rightarrow 3\rightarrow 1$.  In the remainder
of this section we will give explicit expressions associated with the
$X$ variable, with the understanding that the formulas related to $Y$
and $Z$ can be constructed by analogy.

A more accurate approximation to the TDI response is the so-called
rigid adiabatic (RA) approximation~\cite{Rubbo:2003ap}, which is valid
in the regime where the finite lengths of data used to approximate the
idealized Fourier transforms are short enough that the geometry and
orientation of the detector doesn't change significantly during this
time.  In the RA formalism, the response is
\begin{equation}
  \label{e:RAA-R}
  R(f)
  = \frac{ R_{\LWL}(f) \, e^{i4\pi f\Tlight}}{\sinc\left({2\pi f\Tlight}\right)}\,,
\end{equation}
and, for a wave propagating along the unit vector $\khat$,
\begin{equation}
  \label{e:RAA-d}
  \tens{d}^X(f,\khat) =
  \left\{
    \mf{T}_{\nhat_2}(f,\khat)\,\frac{\nhat_2\otimes\nhat_2}{2}
    - \mf{T}_{-\nhat_3}(f,\khat)\,\frac{\nhat_3\otimes\nhat_3}{2}
  \right\}\,,
\end{equation}
where (defining $\xi(\khat)\equiv (1-\khat\cdot\nhat)$)
\begin{equation}
  \mf{T}_{\nhat}(f,\khat)
  =  \frac{e^{i2\pi f\Tlight\khat\cdot\nhat/3}}{2}
\{
    e^{i\pi f\Tlight\xi(\khat)}
    \sinc
[
      \pi f\Tlight\xi(-\khat)
]
    +
    e^{-i\pi f\Tlight\xi(-\khat)}
    \sinc
[
      \pi f\Tlight\xi(\khat)
]
\}
\end{equation}
is a transfer function associated with the arm along $\nhat$.  Note
that this is related to the $\mc{T}_{\nhat}(f,\khat)$ defined
in~\cite{Rubbo:2003ap} by an overall phase, and also that
$\mf{T}_{\nhat}(f,\khat)$ reduces to unity in the LWL
$f\ll 1 / (\pi\Tlight)$.

The input to the LAL/LALApps search code consists of
Fourier-transformed data stretches of duration $T_\SFT$, referred to as
Short Fourier Transforms (SFTs).  This is a common data format used
within the LIGO Scientific Collaboration for continuous-wave searches
(e.g., see~\cite{Abbott:2006vg}).  The time baseline $T_\SFT$
has to be chosen sufficiently short such that the noise-floor can be
approximated as stationary and the rotation and acceleration of the
LISA detector can be neglected, and we chose $T_\SFT = 7$\,days.

We produce ``calibrated SFTs'' by Fourier-transforming the raw TDI data
and applying a
frequency-domain response function to produce a Fourier transformed
strain
\begin{equation}
  \wt{x}^{X}(f) \equiv R(f) \, \wt{X}(f)\,.
\end{equation}
For our MLDC1 analysis~\cite{Prix:2007zh} and MLDC2
submission~\cite{MLDC2Poster} we used the long-wavelength
approximation $R_{\LWL}(f)$ for calibrating SFTs, but for subsequent
analyses (including our MLDC1B search~\cite{Whelan:2008zz}) we have
produced ``rigid adiabatic'' SFTs, which use the full form of $R(f)$
defined in \eqref{e:RAA-R}.

Our pipeline includes modifications to implement the full form of
$\tens{d}^{I}(f,\khat)$.  However, a logistically simpler
intermediate approximation was also used in the initial followup to our
MLDC2 work.  In this ``partial rigid adiabatic'' (pRA) formalism, the more
precise form of $R(f)$ from \eqref{e:RAA-R} is used to
construct the SFTs, but the further analysis proceeds with the simpler
form of $\tens{d}^{I}_{\LWL}$. See \tref{t:RAdefns} for a summary of the
three different levels of response approximation considered in this analysis.
\begin{table}[tbp]
  \label{t:RAdefns}
  \begin{tabular}{||l||c|c|c||}
    \hline
    full name & label & response & detector tensor \\
    \hline
    long-wavelength & LW & $R_{\LWL}(f)$ & $\tens{d}^{I}_{\LWL}$ \\
    partial rigid adiabatic & pRA & $R(f)$ & $\tens{d}^{I}_{\LWL}$ \\
    full rigid adiabatic & RA & $R(f)$ & $\tens{d}^{I}(f,\khat)$ \\
    \hline
  \end{tabular}
  \caption{Definitions of the long-wavelength (LW), partial rigid
    adiabatic (pRA), and full rigid adiabatic (RA) formalisms, in terms
    of the response function $R(f)$ (used to
    calibrate SFTs) and the detector tensor $\tens{d}^{I}(f,\khat)$.}
\end{table}

\section{Search on Mock LISA Data}

\label{s:pipeline}

\subsection{Search Pipeline}

\label{s:pipeline-search}

\begin{figure}
  \begin{center}
    \includegraphics[width=0.6\textwidth,clip]{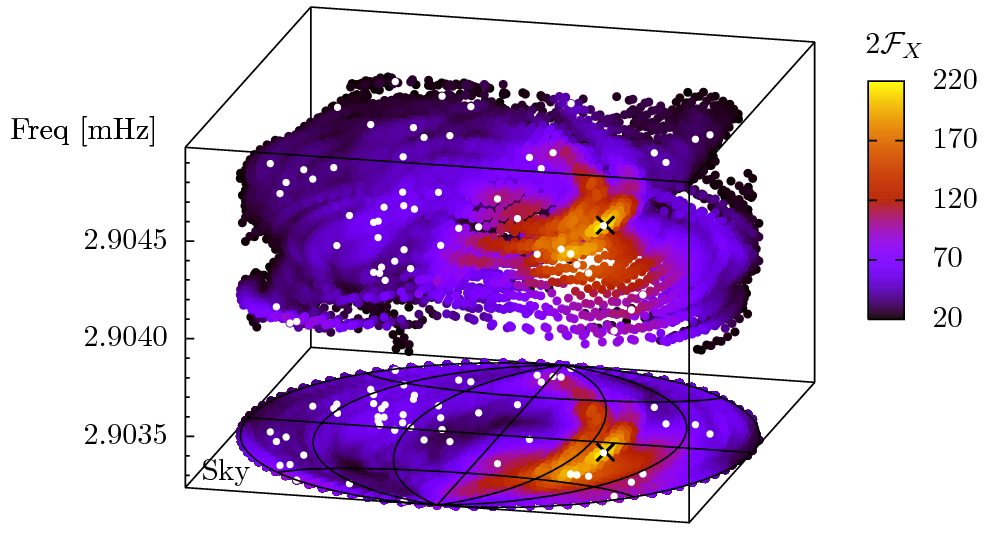}\\
    \includegraphics[width=0.6\textwidth,clip]{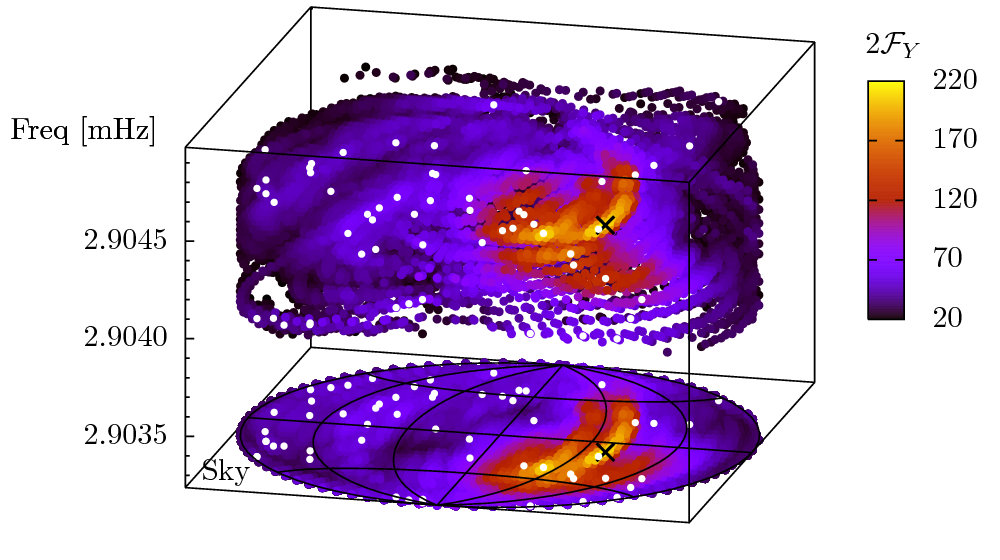}\\
    \includegraphics[width=0.6\textwidth,clip]{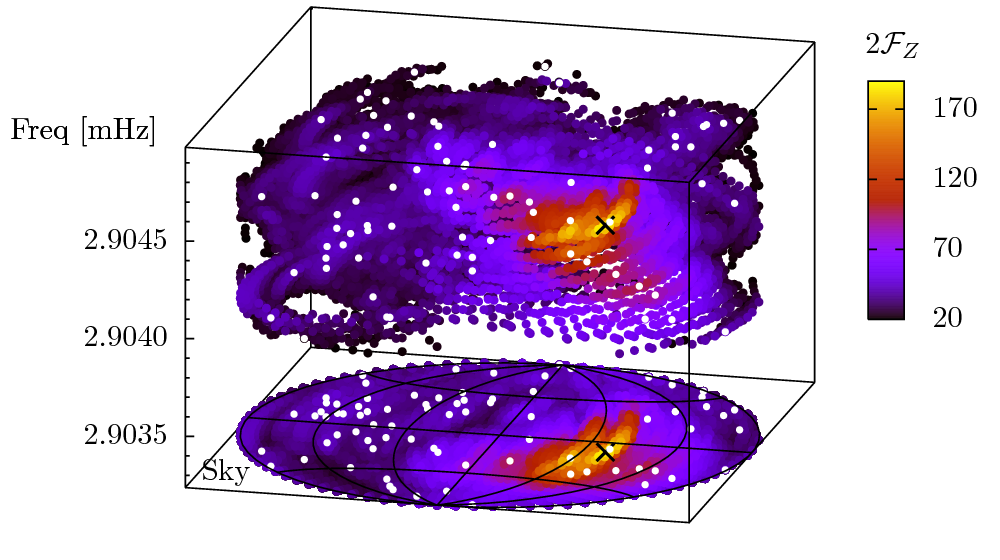}\\
  \end{center}
  \caption
  { Doppler space structure of $\F$-statistic in each of the three TDI variables ($\F_X$,
    $\F_Y$, $\F_Z$) for a single source (indicated by the black cross) at
    $f_{\text{signal}}\sim 2.9044\un{mHz}$. Shown are points with
    $2\F>20$ over the whole sky and within a Frequency window of
    $f_{\text{signal}} \pm 2\times10^{-4}f_{\text{signal}}$.  White circles
    indicate local maxima in $2\F$.  While the absolute maximum at
    the true signal parameters coincides in the three
    ``single-detector'' searches, many secondary maxima are found
    at different points in each of the three searches.  This
    is the basis of the {\coinc} criterion used to distinguish
    between true signals and secondary maxima. }
  \label{fig:paramSpace}
\end{figure}
As in MLDC1~\cite{Prix:2006wm}, we used the standard LAL/LALApps
software~\cite{lalapps} developed for the search for continuous GWs
with ground-based detectors, in particular the code
\texttt{ComputeFStatistic\_v2}, which implements the multi-detector
$\F$-statistic \eqref{e:Fstat}.  We extended our LISA-specific
generalizations of the code to allow analysis in either the
long-wavelength or rigid adiabatic formalisms.

Both single- and multi-detector $\F$-statistic searches are
complicated by the presence of \textit{secondary maxima} in Doppler
parameter space, i.e., points where $\F$ reaches a substantial local
maximum, separated from the \textit{primary maximum} at Doppler
parameters of the true signal.  This is illustrated in
\fref{fig:paramSpace} for a search with only one injected signal.
If only one signal is present, the global maximum of $2\F$ can be
identified as the parameters of the true signal.  Our original MLDC1
search identified the loudest signal within a narrow frequency band as
the true signal, but could not distinguish between secondary maxima
due to that signal and weaker true signals nearby in frequency and at
different points in the sky.

In constructing our MLDC2 pipeline\cite{MLDC2Poster},
we observed empirically that the
same source tended to generate different patterns in secondary maxima
across the sky in the TDI variables $X$, $Y$, and $Z$.  We thus
identified ``true'' signal candidates by requiring them to have
consistent Doppler parameters in single-detector searches performed
using the $X$, $Y$, and $Z$ observables.  (The noise correlation among
those three observables is irrelevant because this stage involves
{\coinc}s among the results of three single-detector searches rather
than a coherent multi-detector search.)
Note that a \emph{coherent} multi-detector search involving $X$, $Y$ and
$Z$ does not have this discriminating power, as it also yields a
likelihood surface with primary and secondary maxima, similar to
\fref{fig:paramSpace}.
The details of the {\coinc}
criterion are discussed below, but it is based on requiring a low
Doppler mismatch~\cite{Prix:2006wm}
\begin{equation}
  \label{e:mismatch}
  m = g_{ij} \delta\doppler^i \delta\doppler^j
  \sim 1 - \frac{E\left[2\F(\doppler + \delta\doppler)\right]}{E\left[2\F(\doppler)\right]}
\end{equation}
between candidates in different single-detector searches.
No condition was placed on the
consistency of the recovered amplitude parameters, in part because the
LW searches in particular are known to produce unreliable Doppler
parameters.
Once signals had been identified in {\coinc} between pairs of
single-detector searches, those candidates were followed up with
finer-gridded single-detector searches, and candidates surviving in
{\coinc} were then targeted with a coherent multi-detector search using the
noise-independent TDI combinations $X$ and $Y-Z$ as the two
``detectors''.

The detailed pipeline was thus as follows:
\begin{itemize}
\item[\textbf{Stage One:}] Wide parameter-space single-detector searches using
    each of the TDI variables $X$, $Y$, and $Z$.  Up to $N=100,000$ Doppler
    parameter points with $2\F>2\F_{\text{th}}=20$ values are kept.  Note
    that only one year of data is used in these initial
    single-detector stages. This was empirically found to cut down on
    false alarms.
\item[\textbf{Stage Two:}] Identification of local maxima in each ``detector''.  A
    signal is a local maximum if there is no signal with higher $2\F$
    value at a Doppler mismatch \eqref{e:mismatch} of
    $m<m_{\text{LM}}=2$.
\item[\textbf{Stage Three:}] Identification of coarse {\coinc}s among the
    searches, using the {\coinc} condition $m<m_{\text{COINC1}}=0.8$.  For
    our search of MLDC2 data, a set of candidates in the three detectors was
    considered to be in {\coinc} if each pair of candidates was within
    the prescribed mismatch.
\item[\textbf{Stage Four:}] Followup of initial candidates with finer-gridded
    single-detector searches and tighter {\coinc}.  The search
    ``zooms in'' by iteratively increasing the resolution of the
    Doppler parameter grid for each detector.  At the end of this
    process, {\coinc} among the detectors is checked again with the
    tighter condition $m<m_{\text{COINC2}}=0.2$.
\item[\textbf{Stage Five:}] Final multi-detector followup of surviving candidates.
    Now a multi-detector search is performed with the TDI combinations
    $X$ and $Y-Z$, which have independent noise contributions.  As in
    stage four, the search ``zooms in'' on the true Doppler parameters
    iteratively.  The ultimate resolution of the search is set by this
    multi-detector search, which uses the full two years of data.
\end{itemize}

\subsection{Parameter Errorbars}

\label{s:pipeline-errorbars}

We estimated the errors expected from Gaussian fluctuations of the noise
using the Fisher information matrices on the amplitude and Doppler
parameter subspaces.  For the amplitude parameters, the expected
discrepancy $\dAmp=\Acand-\Akey$ between the parameters $\Acand$
returned by the search and the true signal parameters $\Akey$ is
described by the expectation value
\begin{equation}
  \label{e:expectdAmp}
  \expected{\dAmp^\mu\dAmp^\nu} = \M^{\mu\nu}(\doppler_\sig)\,,
\end{equation}
so we can quote an errorbar on a particular $\Amp^\mu$ of
\begin{equation}
  \label{e:sigmaAmu}
  \sigma_{\Amp^\mu} = \sqrt{\M^{\mu\mu}}\,,
\end{equation}
with no implied sum over $\mu$ (we made no attempt to translate this
back into errorbars on the physical parameters $h_0$, $\cos\iota$,
$\psi$ and $\phi_0$).

Note, however, that this definition assumes that either the Doppler
parameters $\doppler_\sig$ are perfectly matched by the candidate
(i.e., $\doppler_\cand = \doppler_\sig$) or there are no correlations
between amplitude- and Doppler-coordinates in the full parameter-space
Fisher matrix. In practice none of these two conditions are satisfied
in the present search. Therefore we expect deviations from these
predicted error-distributions even in the case of perfectly Gaussian noise.

For the Doppler parameters, the expected discrepancy $\ddoppler^i =
\doppler^i_\cand - \doppler^i_\sig$, between the Doppler parameters
$\doppler^i_\cand$ of the primary $2\F$ maximum and those
$\doppler^i_\cand$ of the simulated signal is described by
\begin{equation}
  \label{e:expectddop}
  \expected{\ddoppler^i \ddoppler^j} = \Fisher^{ij}\,,
\end{equation}
where $\Fisher^{ij}$ is the inverse of the Fisher information matrix
\begin{equation}
  \Fisher_{ij} = g_{ij}\,\abs{\Acand}^2\,,
\end{equation}
which can be defined in terms of the Doppler metric $g_{ij}$ associated with
the mismatch \eqref{e:mismatch}. Similar to the error-estimates on
amplitudes, this definition assumes either perfectly matched amplitude
parameters (i.e., $\Acand = \Akey$) or a block-diagonal Fisher matrix
over the full parameter space, with no correlations between
amplitude- and Doppler-space. In practice none of these conditions is
true and we therefore expect deviations from the predicted error estimates.

Rather than the full $\F$-statistic metric, we use the approximate
\emph{orbital metric}~\cite{Prix:2006wm}.  This metric is approximated
having constant elements in terms of the {\coord}s
$\{\omega_0,k_x,k_y\}$ where
\begin{subequations}
  \begin{align}
    \omega_0 &= 2\pi f \,,\\
    k_x &= - 2\pi f \frac{v_{\text{orb}}}{c} \cos\lat\cos\lon\,, \\
    k_y &= - 2\pi f \frac{v_{\text{orb}}}{c} \cos\lat\sin\lon\,.
  \end{align}
\end{subequations}
A limitation of the orbital metric is that it cannot distinguish
between the points $\{f,\lat,\lon\}$ and $\{f,-\lat,\lon\}$
which are reflected through the ecliptic.  (The search itself can
distinguish between points with different signs of ecliptic latitude,
thanks to the different amplitude modulation, but this is not captured
in the orbital metric.)  Assigning error bars to the frequency $f$ and
ecliptic longitude $\lon$ is straightforward, but converting an
uncertainty in $\cos\lat$ into an uncertainty in ecliptic latitude
$\lat$ is complicated by the orbital metric becoming singular at the
ecliptic.  As a workaround, we first calculate error bars
\begin{subequations}
  \begin{align}
    \label{e:sigmaFreq}
    \sigma_f^2 &= \frac{\partial f}{\partial\doppler^i}
    \frac{\partial f}{\partial\doppler^j} \Fisher^{ij}\,, \\
    \label{e:sigmaLon}
    \sigma_\lon^2 &= \frac{\partial\lon}{\partial\doppler^i}
    \frac{\partial\lon}{\partial\doppler^j} \Fisher^{ij} \,,\\
    \sigma_{\cos\lat}^2 &= \frac{\partial\cos\lat}{\partial\doppler^i}
    \frac{\partial\cos\lat}{\partial\doppler^j} \Fisher^{ij}\,,
  \end{align}
  and then estimate the error in $\lat$ as
  \begin{equation}
    \label{e:sigmaLat}
    \sigma_\lat := \cos^{-1}(\cos\lat - \sigma_{\cos\lat}) - \abs{\lat}\,,
  \end{equation}
  so that
  \begin{equation}
    cos(\abs{\lat} + \sigma_\lat) = \cos\lat + \sigma_{\cos\lat}\,,
  \end{equation}
  i.e., we match the one-sigma equation in the direction away from the
  ecliptic, where the conversion between $\lat$ and $\cos\lat$ should
  be well behaved.  It is of course possible that
  $\sigma_\lat>\abs{\lat}$, in which case the $\pm1\sigma$ interval we
  define straddles the ecliptic, but this agrees qualitatively with
  the observation that some signals near the ecliptic are recovered in
  the opposite hemisphere.
\end{subequations}

\section{Evaluation}

\label{s:eval}

\subsection{Signal Identification}

\label{ss:eval-id}

When run on the MLDC 2.1 dataset, our pipeline returns $\sim 2000$
signals found in {\coinc}.  To evaluate its performance, we check how
many of those sources were found at parameters consistent with those
of one of the galactic binary signals injected into the data.  The
original datasets were generated with $\sim 30\,$million signals, but
of those {\numBright} were considered ``bright'' enough to detect by
the MLDC Task Force and their parameters were placed into a separate
key file.  It is against that key that we compare our results.

In part due to the known inaccuracies in amplitude parameters
associated with the long-wavelength and partial rigid adiabatic
responses, we checked for consistency using only the Doppler
parameters (frequency and sky position).  A signal was considered to
be ``found'' if the Doppler parameters of the candidate and the key
had a mismatch
\begin{equation}
  m_{\cand\sig} = g_{ij} \ddoppler^i \ddoppler^j
\end{equation}
of 1 or less.  (In the case of multiple injected and/or candidate
signals satisfying the mismatch condition, the brightest were ``paired
off'' first.)  If no injected signal matched a candidate, that
candidate was considered to be a ``false alarm''.

\subsection{Parameter Errors}

\label{ss:eval-errors}

These expectation values \eqref{e:expectdAmp} and \eqref{e:expectddop}
allow us to define, as in \cite{Whelan:2008zz},
\begin{equation}
  \label{e:dopError}
  \dopErr = \sqrt{\frac{\Fisher_{ij}}{3} \ddoppler^i \ddoppler^j}\,,
\end{equation}
and
\begin{equation}
  \label{e:ampError}
  \ampErr = \sqrt{\frac{\M_{\mu\nu}}{4} \dAmp^\mu \dAmp^\nu}
  = \frac{\abs{\dAmp}}{2}\,,
\end{equation}
so that
\begin{equation}
  \expected{\dopErr^2} = 1 = \expected{\ampErr^2}\,,
\end{equation}
in the ideal case of statistical errors due to Gaussian noise, and no
correlations between amplitude- and Doppler-parameters.

\section{Results without signal subtraction}

\label{s:results}

\subsection{Signal Recovery}

The signal recovery of our pipeline using the various response models
is summarized in \tref{t:compTable}.  Note that signal recovery
is not significantly affected by the scalar response function $R(f)$,
so the pRA search performs almost identically to the LW one.
(See \sref{ss:method-resp} and \tref{t:RAdefns} for the definitions of
the different models of the LISA response.)  However,
the use of the full response tensor $\tens{d}(f,\khat)$ in the full RA
search leads to an increase in the number of found signals from
{\numFoundLWL} to {\numFound}, with much of the improvement coming from
higher-frequency signals, and a reduction in the number of false
alarms.  Note that while the designated ``bright signals'' key
contained {\numBright} sources, many of those were still not bright by
the standards of our search, having low values of $\abs{\Akey}^2$.
Since the faintest signal found by our search had
$2\F=\abs{\Acand}^2=\twoFMin$, we focus attention on sources with
$\abs{\Akey}^2\ge 40$, of which there are {\numForty} in the key file.
\begin{table}[tbp]
  \label{t:compTable}
  \begin{tabular}{||c||c||c|c|c||c|c|c||}
\hline
\multirow{2}{*}{Freqs} & Signals & \multicolumn{3}{c||}{Found} & \multicolumn{3}{c||}{False}
\\
& ($\abs{\Amp}^2>40$) & RA & pRA & LW & RA & pRA & LW
\\
\hline
\hline
0--5\,mHz& 4446& 1025& 984& 982& 2& 1& 1\\
\hline
5--10\,mHz& 1967& 822& 652& 652& 3& 5& 5\\
\hline
10--15\,mHz& 163& 133& 68& 68& 0& 1& 1\\
\hline
15--20\,mHz& 7& 7& 2& 2& 0& 0& 0\\
\hline
20--27\,mHz& 3& 2& 0& 0& 0& 2& 2\\
\hline
\hline
Total& 6586& 1989& 1706& 1704& 5& 9& 9\\
\hline
\end{tabular}

  \caption{
    Comparison of galactic WDB signal recovery with long-wavelength (LW),
    partial rigid adiabatic (pRA) and full rigid adiabatic (RA) response.
    The pRA and LW response functions differ only in the scalar piece
    $R(f)$, and produce similarly efficient searches.  The RA response,
    including the full response tensor $\tens{d}(f,\khat)$, leads to
    more found signals and fewer false alarms, especially at higher
    frequencies.
  }
\end{table}

\subsection{Doppler parameter accuracy}

\label{ss:results-dopErrors}

\begin{figure}
  \begin{center}
    \includegraphics[width=0.8\textwidth]{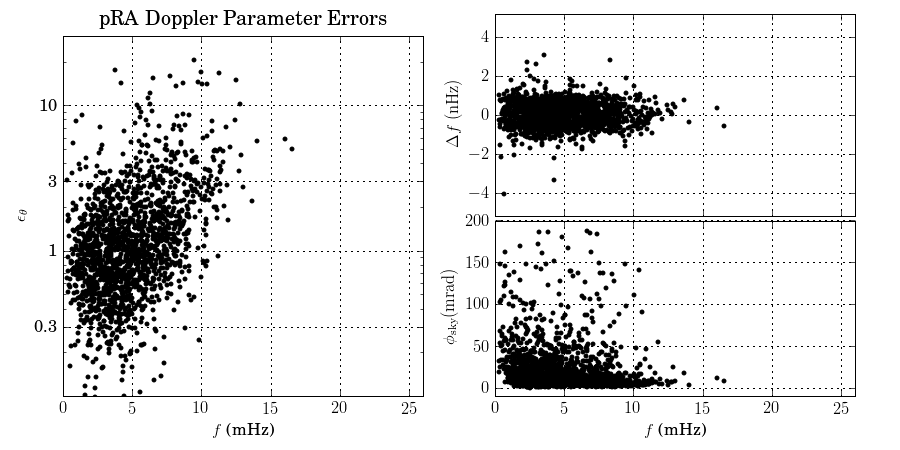}
    \includegraphics[width=0.8\textwidth]{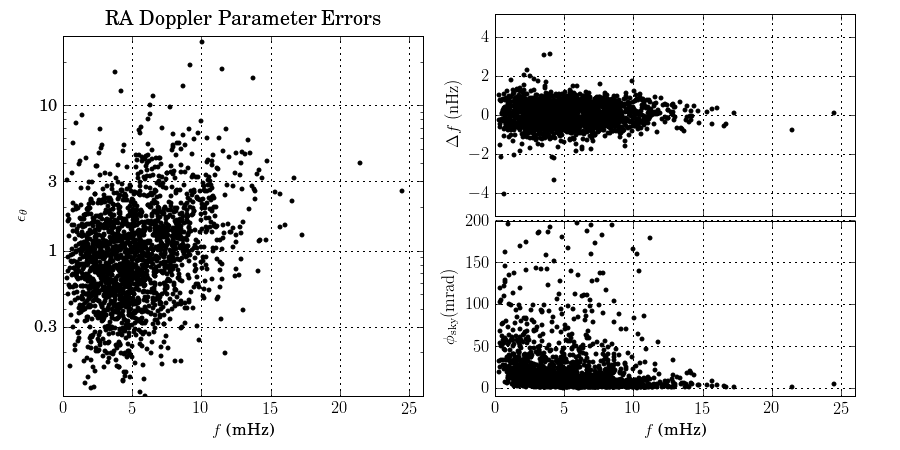}
  \end{center}
  \caption
  { Doppler parameter errors as a function of frequency.  The left
    panel shows the error measure $\dopErr$ defined in
    \eqref{e:dopError} which is normalized to have expected unit RMS
    value in the case of perfectly matched amplitude parameters
    and statistical errors caused by Gaussian noise.
    The right panels show the errors in frequency and sky position.
    We show the errors for searches using the partial rigid adiabatic
    (pRA) and full rigid adiabatic (RA) responses.  The Doppler
    errors using the na\"{\i}ve long-wavelength limit are nearly
    identical to the pRA results and are therefore omitted.
  }
  \label{f:dopErrors}
\end{figure}

\begin{figure}
  \begin{center}
    \includegraphics[width=0.8\textwidth]{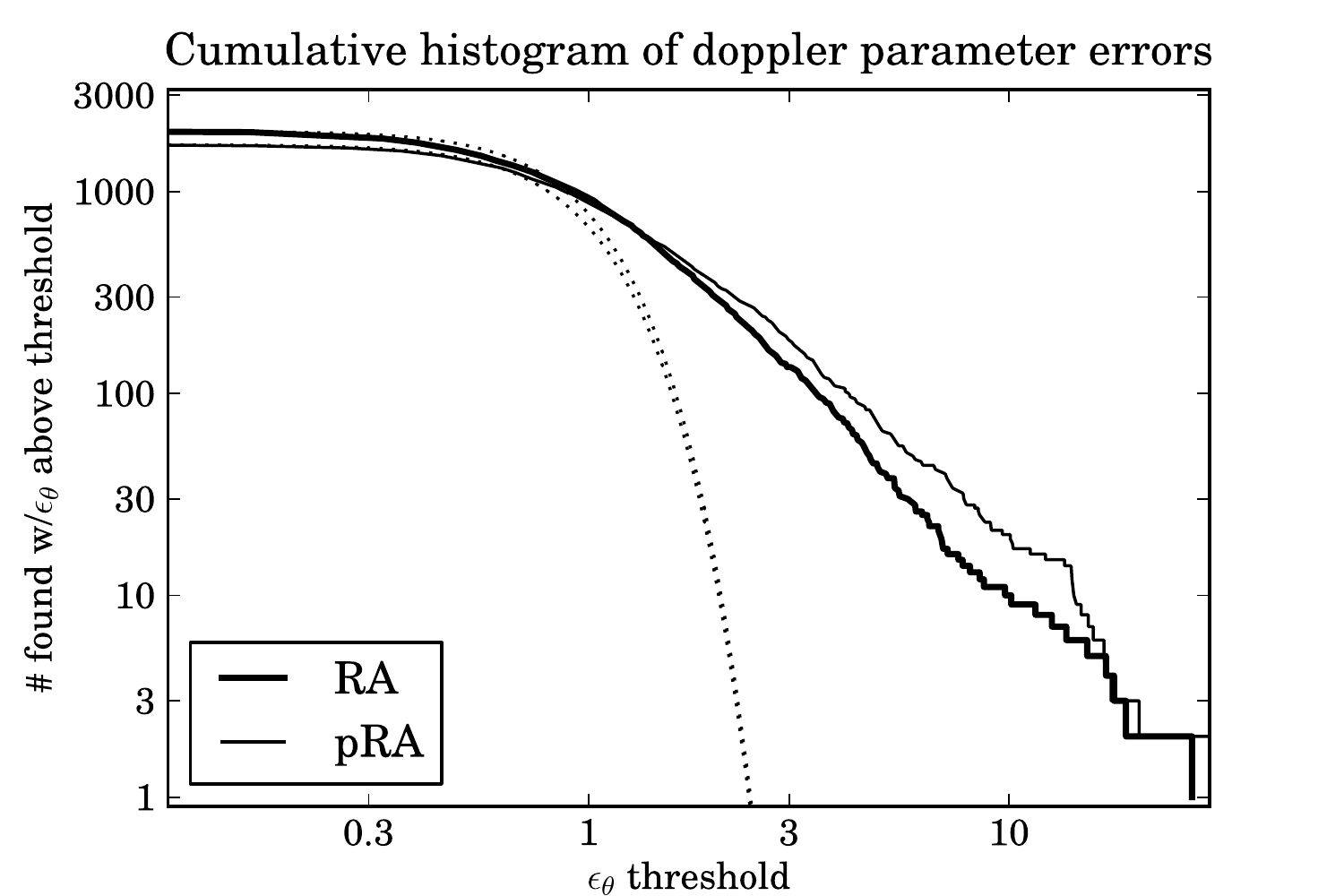}
  \end{center}
  \caption
  { Cumulative histograms for Doppler parameter errors $\dopErr$ (see
    \fref{f:dopErrors} for definitions).  The dotted lines show
    the appropriately scaled theoretical CDF \eqref{e:dopGauss} in the
    case of perfectly matched amplitude parameters and statistical
    errors caused by Gaussian noise, where $3\dopErr^2$ follows a
    central $\chi^2$ distribution with three degrees of freedom.
  }
  \label{f:dopHist}
\end{figure}

As described in \sref{ss:eval-errors}, we can compare the Doppler parameters
$\doppler_\cand$ of each candidate with those $\doppler_\sig$ of the
corresponding signal in the key.  The errors, as a function of
candidate frequency, are shown in \fref{f:dopErrors}.  We show the
errors for the searches using the full (RA) and partial (pRA) rigid
adiabatic responses.  The Doppler parameters returned using the
long-wavelength (LW) limit are almost identical to the pRA results, so
we omit those.  We plot the combined measure $\dopErr$ defined in
\eqref{e:dopError} (which has $\expected{\dopErr^2}=1$ in the case of
Gaussian statistical errors), the error $\Delta f$ in the measured
frequency and the angle $\phi_{\text{sky}}$ between the recovered and
actual sky positions.
The pRA Doppler errors are seen to be a bit larger than the RA Doppler
errors.  We can quantify this by making a cumulative histogram of the
Doppler error measure $\dopErr$ defined in \eqref{e:dopError}.  For
statistical errors arising from Gaussian noise and neglecting
amplitude-Doppler correlations, $3\dopErr^2$ should
follow a central $\chi^2$ distribution with three degrees of freedom,
i.e.,
\begin{equation}
  \label{e:dopGauss}
  P(\dopErr>\dopErr^*) = \erfc\left(\dopErr^*\sqrt{\frac{3}{2}}\right)
    + \sqrt{\frac{6}{\pi}}\,\dopErr^*\,e^{-3{\dopErr^*}^2/2}\,.
\end{equation}
The cumulative histograms of $\dopErr$ for the RA and pRA search,
together with these theoretical expectations, are plotted in
\fref{f:dopHist}.

\subsection{Amplitude parameter accuracy}

\label{ss:results-ampErrors}

As described in \sref{ss:eval-errors}, we compare the vector of Amplitude
parameters $\Acand$ of each candidate with those $\Akey$ of the
corresponding signal in the key.  The natural geometrical structure
for studying those vectors is the metric $\mc{M}_{\mu\nu}$.  (For
concreteness we use $\mc{M}_{\mu\nu}(\doppler_\sig)$, i.e.\ evaluated at the
Doppler parameters of the signal in the key).  We can use the quantity
$\ampErr$ defined in \eqref{e:ampError} (which has
$\expected{\ampErr^2}=1$ in the case of Gaussian statistical errors
and perfectly matched Doppler parameters)
as a measure of the overall discrepancy.  We can also separate out
the discrepancy in the length of the amplitude vectors from the angle
between them in the four-dimensional amplitude parameter space, using
quantities defined in \cite{Whelan:2008zz}:
\begin{equation}
  \label{e:delA}
  \delA \equiv \frac{\abs{\Acand}^2 - \abs{\Akey}^2 - 4}{2 \,\abs{\Akey}^2}
\end{equation}
is a measure of the amplitude discrepancy designed to have
$\expected{\delA}=0$, while
\begin{equation}
  \label{e:phiA}
  \phiA \equiv \cos^{-1}
  \left(
    \frac{\Acand^\mu\M_{\mu\nu}\Akey^\nu}{\abs{\Acand}\abs{\Akey}}
  \right)
\end{equation}
measures the angle between the amplitude parameter vectors, and can be
thought of as a phase discrepancy.  Both $\delA$ and $\phiA$ have
expected standard deviation $\abs{\Akey}^{-1}$ in the case of Gaussian
statistical errors.

\begin{figure}
  \begin{center}
    \includegraphics[width=0.8\textwidth]{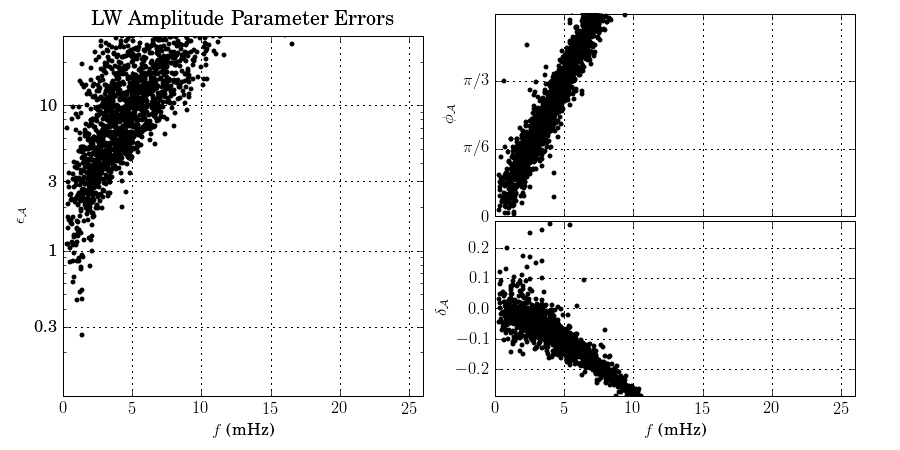}
  \end{center}
  \caption
  { Amplitude parameter errors using the long-wavelength limit (LWL)
    approximation, as a function of frequency.
The left panel shows the error measure $\ampErr$
    defined in \eqref{e:ampError} which is normalized to have expected
    unit RMS value in the case of statistical errors caused by
    Gaussian noise.  The right panels show the errors in the magnitude
    and direction of the amplitude parameter vector, as measured by
    the amplitude difference $\delA$ defined in \eqref{e:delA} and the
    angle $\phiA$ between the true and recovered amplitude parameter
    vectors, defined in \eqref{e:phiA}  Note the frequency-dependent
    systematic phase error $\phiA$ and loss in SNR ($\delA<0$),
    discussed in \sref{ss:results-ampErrors}. }
  \label{f:LWL_LWL_ampErrors}
\end{figure}

\begin{figure}
  \begin{center}
    \includegraphics[width=0.8\textwidth]{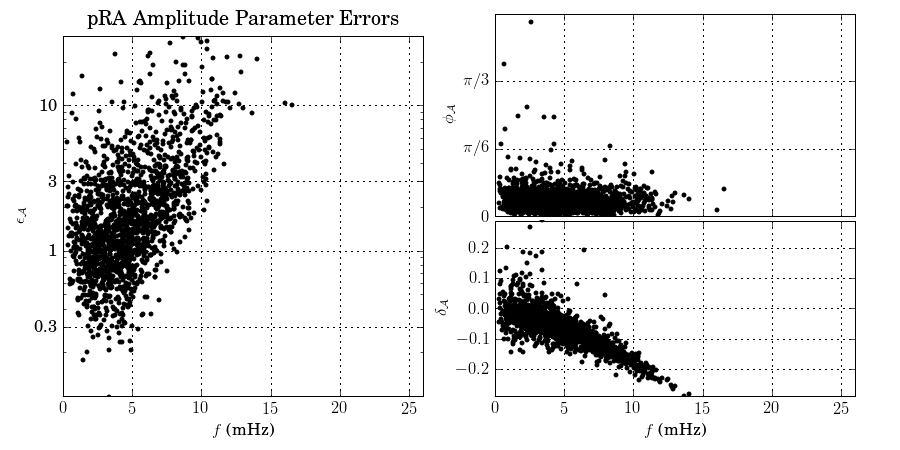}
  \end{center}
  \caption
  { Amplitude parameter errors using the partial rigid adiabatic (pRA)
    approximation, as a function of frequency.  The quantities shown
    are as in \fref{f:LWL_LWL_ampErrors}.  Note the frequency-dependent
    systematic loss in SNR ($\delA<0$), discussed in
    \sref{ss:results-ampErrors}. }
  \label{f:RA_LWL_ampErrors}
\end{figure}

\begin{figure}
  \begin{center}
    \includegraphics[width=0.8\textwidth]{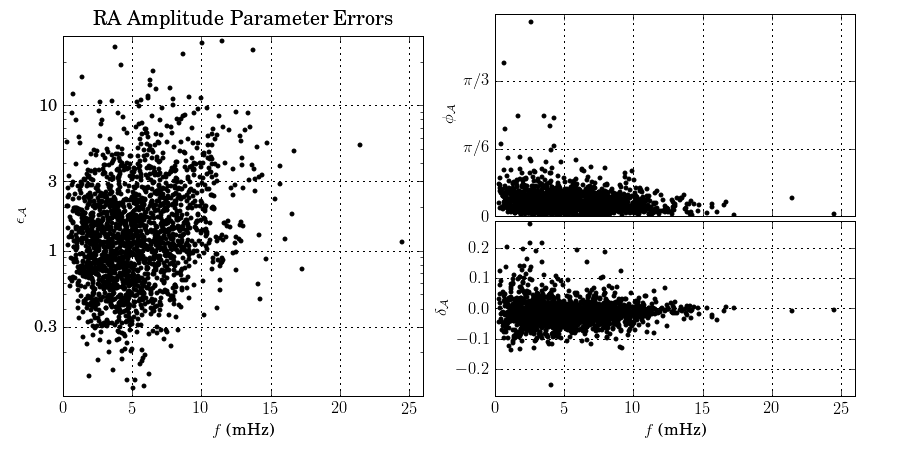}
  \end{center}
  \caption
  { Amplitude parameter errors using the full rigid adiabatic (RA)
    approximation, as a function of frequency.  The quantities shown
    are as in \fref{f:LWL_LWL_ampErrors}.  Note that the
    frequency-dependent systematic phase error seen in
    \fref{f:LWL_LWL_ampErrors} and loss in SNR seen in
    \fref{f:LWL_LWL_ampErrors} and \fref{f:LWL_LWL_ampErrors} are
    absent, as discussed in \sref{ss:results-ampErrors}.  }
  \label{f:RA_RAA_ampErrors}
\end{figure}

In \fref{f:LWL_LWL_ampErrors}, \fref{f:RA_LWL_ampErrors}, and
\fref{f:RA_RAA_ampErrors} we plot $\ampErr$, $\delA$ and $\phiA$ for
the signals recovered in the LWL, pRA, and RA searches, respectively.
The LWL results have substantial systematic errors in both $\delA$ and
$\phiA$.  The error in $\phiA$, which increases linearly with
frequency, turns out to be mostly due to a systematic error in the
initial phase $\phi_0$ corresponding to a time shift of $2\Tlight$.
This problem can be traced to the absence of the factor of $e^{i4\pi
  f\Tlight}$ from \eqref{e:RAA-R} in the scalar response function
$R_{\LWL}(f)$ used in the LW search, and is fixed in the partial rigid
adiabatic (pRA) approximation.  There is also a systematic trend
towards negative $\delA$, which, recalling that
$2\F=\abs{\Acand}^2$, corresponds to a signal being recovered with
lower SNR than expected from the true amplitude parameters.  Part of
this effect is removed by the inclusion of
$\sinc\left({2\pi f\Tlight}\right)$ in the denominator of
\eqref{e:RAA-R}, but the pRA results still show show a
frequency-dependent SNR deficiency.  Both effects are absent in the
full RA search.

\begin{figure}
  \begin{center}
    \includegraphics[width=0.8\textwidth]{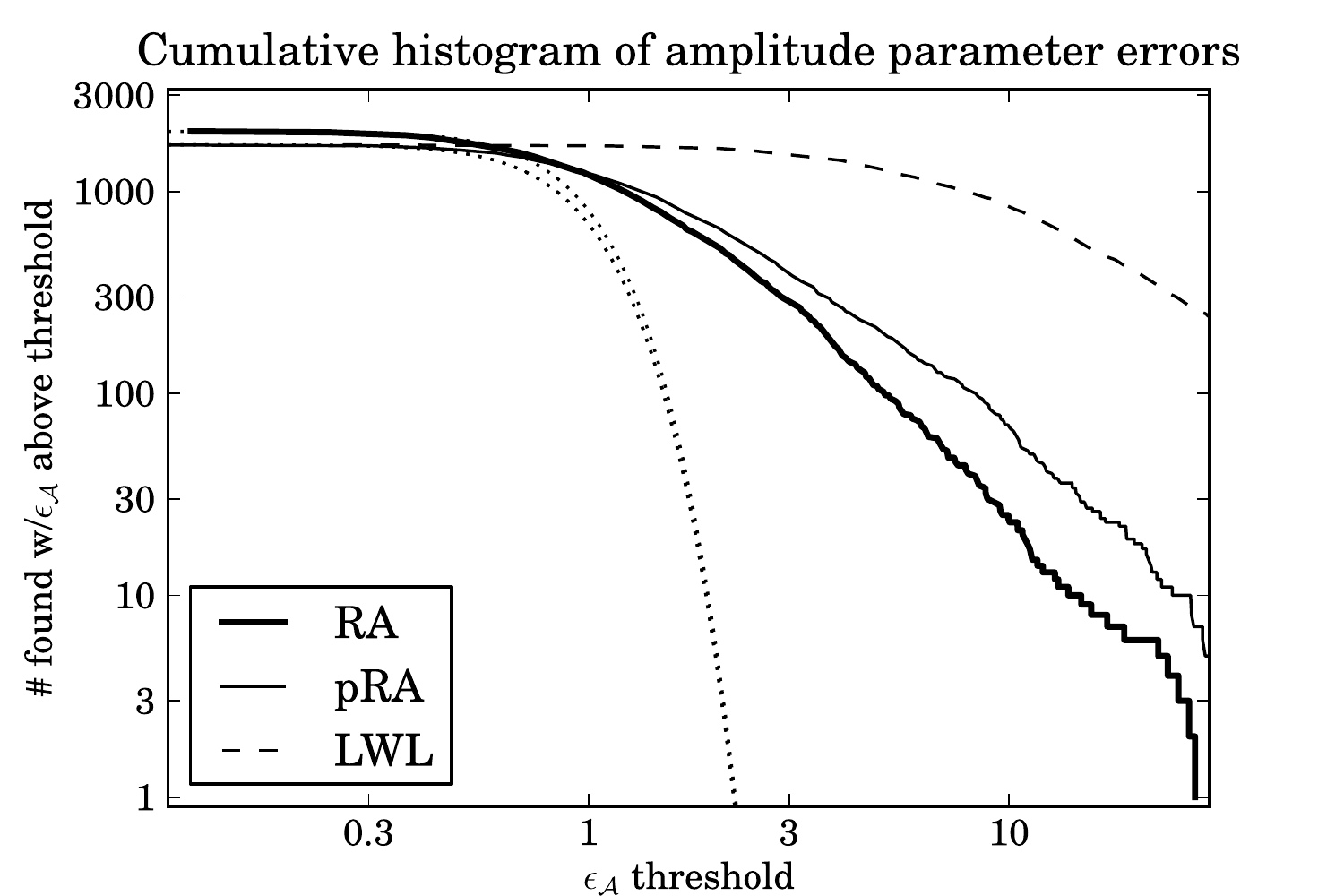}
  \end{center}
  \caption
  { Cumulative histograms for Amplitude parameter errors $\ampErr$
    (see \fref{f:RA_RAA_ampErrors} for definitions).  The dotted lines show
    the appropriately scaled theoretical CDF \eqref{e:ampGauss} in the
    case of statistical errors caused by Gaussian noise (and
    neglecting amplitude-Doppler correlations), where
    $4\ampErr^2$ follows a central $\chi^2$ distribution with four
    degrees of freedom. The systematic errors illustrated by the
    non-Gaussian tails are greatly reduced by using the rigid
    adiabatic (RA) response. }
  \label{f:ampHist}
\end{figure}

We can quantify the systematic errors, especially those still present
in the RA search, by making a cumulative histogram of $\ampErr$.  For
statistical errors arising from Gaussian noise (and neglecting
amplitude-Doppler correlations), $4\dopErr^2$ should
follow a central $\chi^2$ distribution with four degrees of freedom,
i.e.,
\begin{equation}
  \label{e:ampGauss}
  P(\ampErr>\ampErr^*) = e^{-2{\ampErr^*}^2} \left(1+2{\ampErr^*}^2\right)\,.
\end{equation}
The cumulative histograms of $\ampErr$ for the RA, pRA, and LWL
searches, together with these theoretical expectations, are plotted in
\fref{f:ampHist}.

\subsection[Errors relative to errorbars]{Errors relative to estimated errorbars}

\begin{figure}
  \begin{center}
    \includegraphics[width=0.8\textwidth]{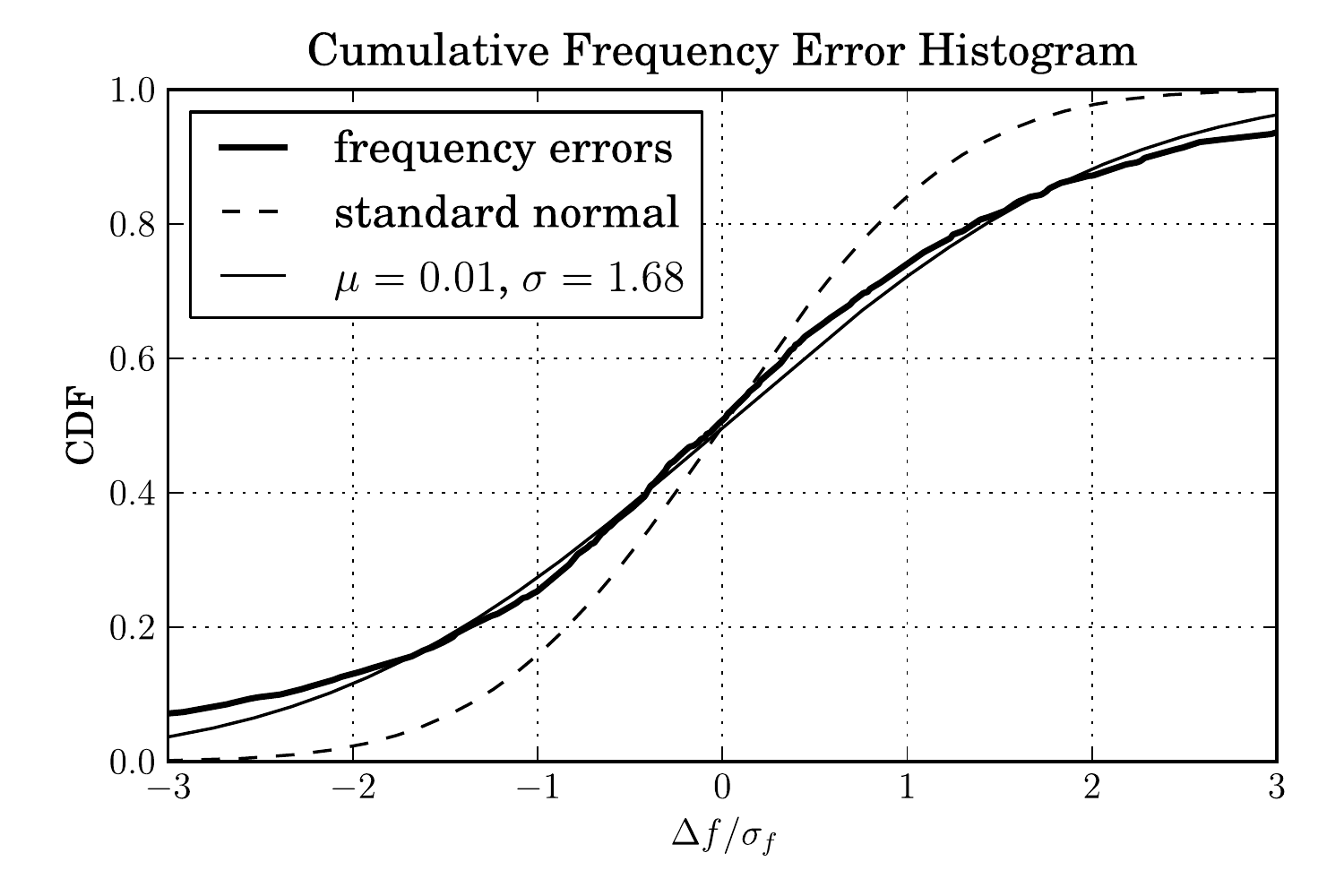}
  \end{center}
  \caption
  { Cumulative histogram of frequency errors as a fraction of the
    errorbars estimated according to \eqref{e:sigmaFreq}, for the full
    RA search.  We plot the histogram of the actual values, along with
    those for two Gaussian distributions: one with unit variance and
    zero mean (standard normal) and one with a mean of $\RAvRAAvmuF$
    and standard deviation of $\RAvRAAvsigF$, equal to those in the
    actual distribution of $\Delta{f}/\sigma_f$ values. }
  \label{f:sigHistFreq}
\end{figure}

\begin{figure}
  \begin{center}
    \includegraphics[width=0.8\textwidth]{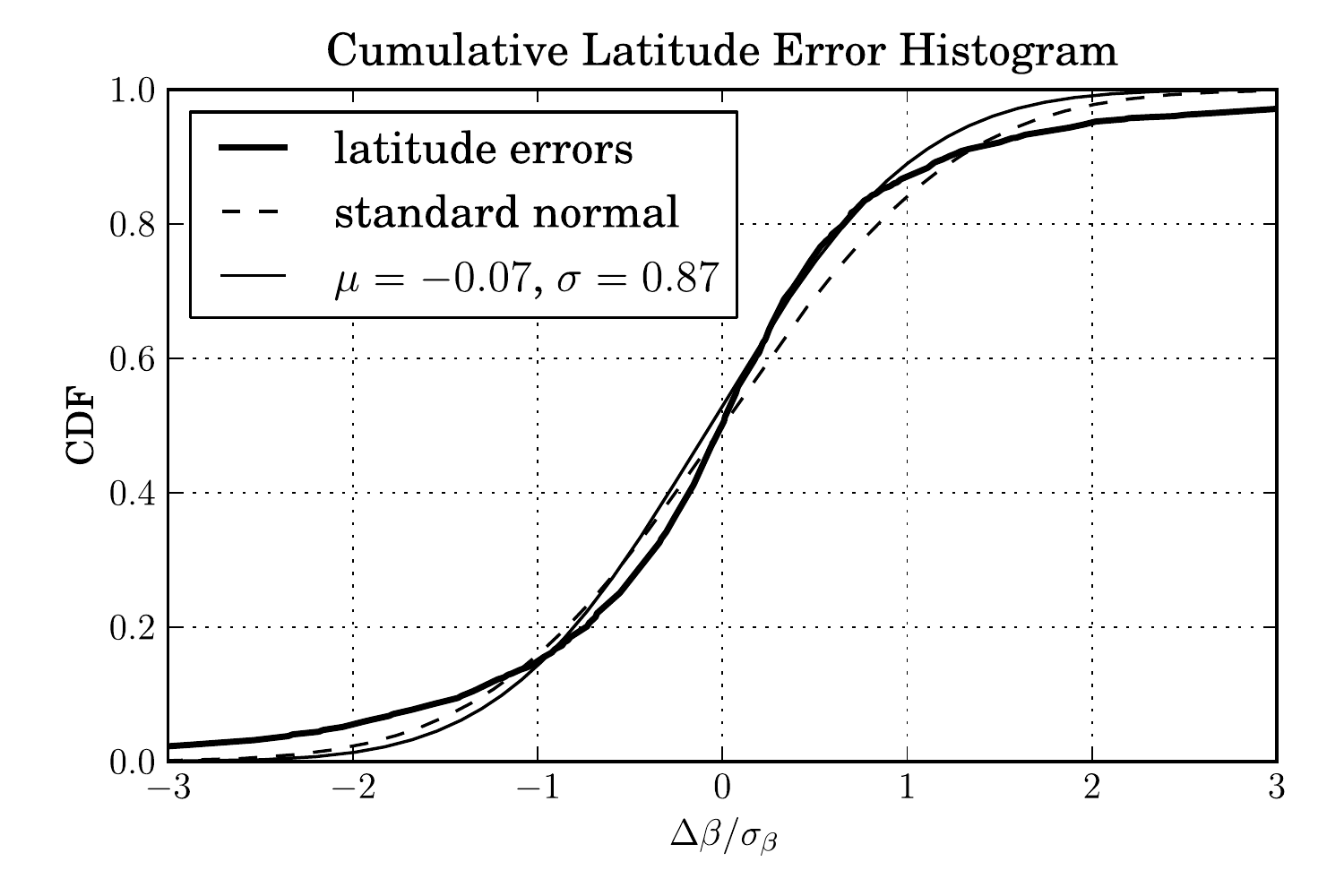}
    \includegraphics[width=0.8\textwidth]{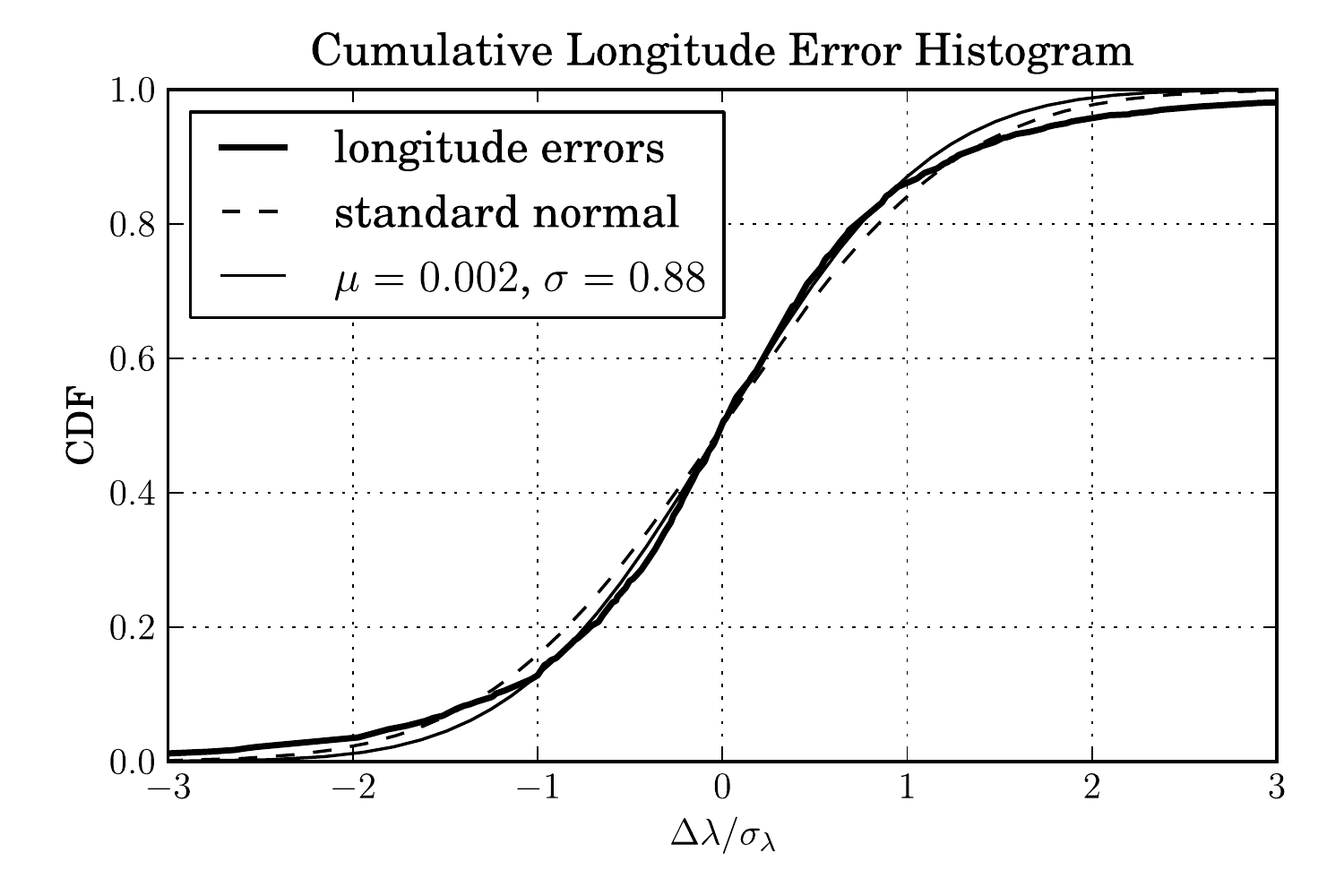}
  \end{center}
  \caption
  { Cumulative histograms for errors in the recovered galactic
    latitude $\lat$ and longitude $\lon$ as a fraction of the
    errorbars estimated according to \eqref{e:sigmaLat} and
    \eqref{e:sigmaLon}, for the full RA search, along with Gaussian
    distributions for comparison as in \fref{f:sigHistFreq}. }
  \label{f:sigHistLatLon}
\end{figure}

Another quantitative comparison of the size of the parameter errors to
theoretical expectations is the ratio of the actual parameter errors
$\dAmp^\mu$ or $\ddoppler^i$ to the errorbars $\sigma_{\Amp^\mu}$ or
$\sigma_{\doppler^i}$ defined in \sref{s:pipeline-errorbars}.  We plot
these for the full rigid adiabatic search.  In \fref{f:sigHistFreq} we
histogram the relative frequency errors $\Delta{f}/\sigma_f$ for the
{\RAvRAAvnumSigs} recovered signals.  For comparison, we also plot the
cumulative histogram for a standard normal distribution, and for one
with a mean $\RAvRAAvmuF$ and standard deviation of $\RAvRAAvsigF$,
the values measured from the actual $\Delta{f}/\sigma_f$ distribution.
In \fref{f:sigHistLatLon} we make similar histograms of the errors in
the galactic latitude $\lat$ and longitude $\lon$.  We see that while
the errors in sky location are comparable in scale to the computed
errorbars, the frequency errors are slightly larger than expected,
with a standard deviation of {\RAvRAAvsigF} (rather than unity) on
$\Delta{f}/\sigma_f$.  (This error is of comparable size in the RA,
pRA, and LW searches.)
For the amplitude parameters, we collect together
$\{\dAmp^\mu/\sigma_{\Amp^\mu}|\mu=1,2,3,4\}$ for each of the
{\RAvRAAvnumSigs} recovered signals, and histogram those
{\RAvRAAvnumAmuSigs} values.  We see a slight systematic excess in
these errors, with a standard deviation of {\RAvRAAvsigAmu} on the
distribution of $\dAmp^\mu/\sigma_{\Amp^\mu}$ values.  This is smaller
than the corresponding value of {\RAvLWLvsigAmu} from the pRA search
and much smaller than the {\LWLvLWLvsigAmu} found in the LW search.
Replacing the long-wavelength approximation with the full rigid
adiabatic response gives us amplitude parameters we can trust.
\begin{figure}
  \begin{center}
    \includegraphics[width=0.8\textwidth]{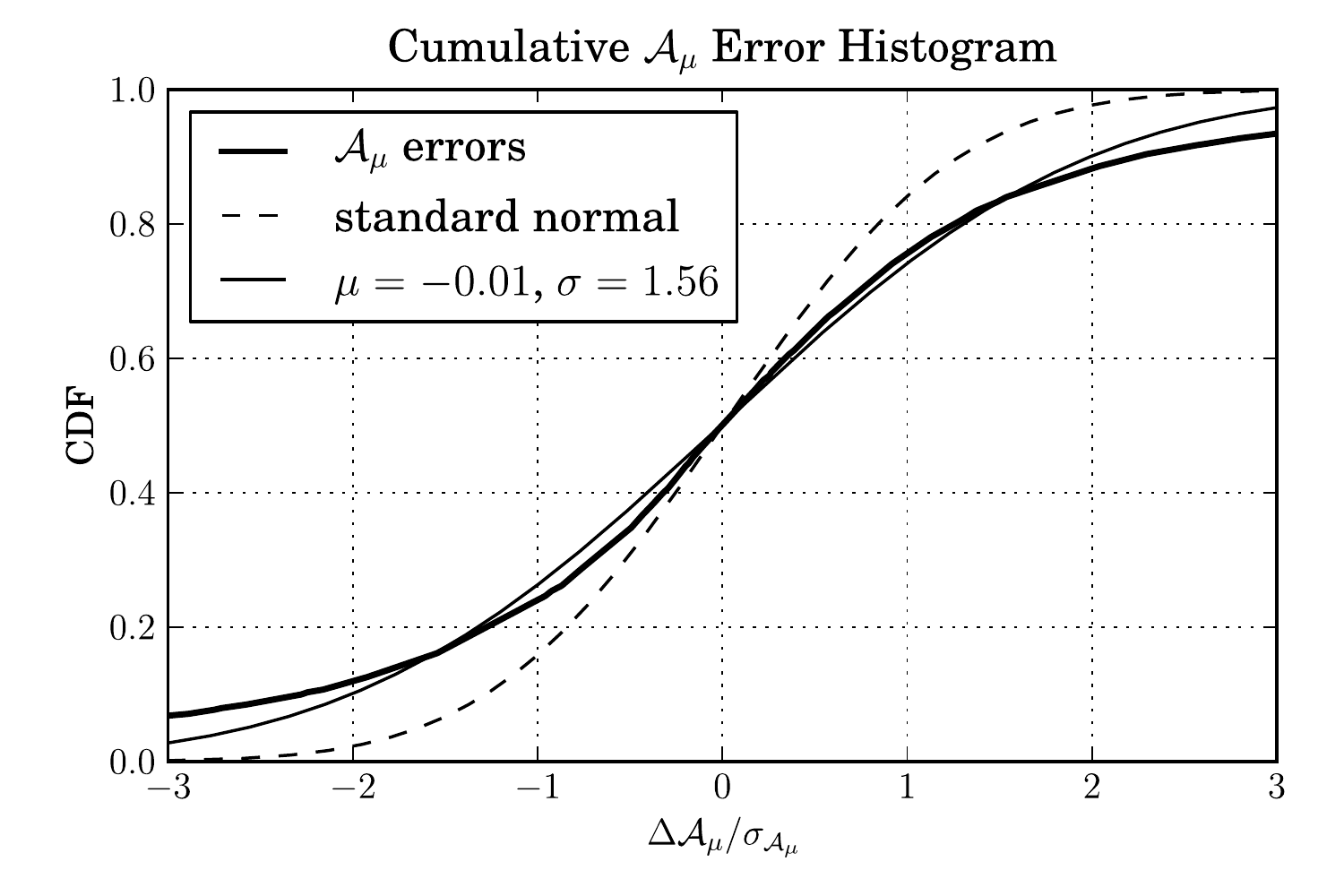}
  \end{center}
  \caption
  { Cumulative histogram of the errors in all amplitude parameters
    $\Amp^\mu$ a fraction of the errorbars estimated according to
    \eqref{e:sigmaAmu}, for the full RA search, along with Gaussian
    distributions for comparison as in \fref{f:sigHistLatLon}. }
  \label{f:sigHistAmu}
\end{figure}

\section{Results with signal subtraction}

\label{s:subtract}

A known limitation of our search pipeline is that it identifies
individual signals, treating all of the other signals as background
noise.  One proposed approach is to generate signals corresponding to
the candidates returned by a search of the data, subtract those from
the data, and re-run the search on the resulting residuals.  (This
approach is a pedestrian alternative to the multiple-signal templates
used in \cite{Crowder:2006eu,Crowder:2007ft}.)
This approach is only likely to work
in a search which returns reliable signal parameters, and so we did
not attempt it with our original long-wavelength-approximation
pipeline.  However, since the pipeline using the rigid adiabatic
response generates very few false alarms, and reasonable Doppler and
amplitude parameter accuracy, we can use it for a simple, illustrative
signal subtraction program.  The algorithm we use is as follows:
\begin{enumerate}
\item Run the original dataset through the standard pipeline described
  in \sref{s:pipeline} to obtain a set of candidate signals.
\item Invert the amplitude parameter vectors ($\Amp^\mu\longrightarrow
  -\Amp^\mu$) of the candidate signals, and use those parameters to
  generate a composite signal to cancel out the signals found so far.
  For this step we used the lisatools~\cite{lisatools} routines used to
  generate the original challenge data, in particular the FastGalaxy
  code~\cite{Cornish:2007if} and synthetic
  LISA~\cite{synthLISA}\footnote{This convenient use of existing
    infrastructure had the potential drawback that differences between
    these signal-generation algorithms and the signal models used in
    our search could lead to imperfections in the cancellation of
    signals.}.
\item Add this cancellation data set to the original MLDC
  data and generate a new set of SFTs with the found signals
  subtracted out.
\item Run the signal-subtracted dataset through the standard pipeline
  to obtain a set of ``new'' candidate signals.
\item Compare the new signals with the signals already found to
  distinguish duplicates (corresponding to unmatched residual portions
  of signals) from truly new signals.  This is done using the same
  matching criterion used to compare found signals with a key,
  described in \sref{ss:eval-id}.  In this step, duplicates are
  analogous to ``matched'' signals and truly new signals are analogous
  to ``false alarms''.
\item Combine the old and new signal lists to obtain a master list of
  signals found so far.  Duplicates are only listed once, using the
  Doppler parameters with which they were found in the original
  search, and combined amplitude parameter vectors
  $\Amp^\mu_{\text{old}}+\Amp^\mu_{\text{new}}$.
\item Repeat the process from step 2, using the new master list of signals
  found so far to subtract from the original dataset.
\end{enumerate}
The results of the iterative procedure are shown in
\tref{t:subtraction}.  We performed {\subtRounds} rounds of iterative
subtraction and re-analysis (stopping the process at that point
because the last round only added two new signals).  This procedure
increased the number of ``true'' signals found from {\numFound} to
{\subtNumFound}, but also increased the false alarm rate, with the
number of false alarms going from {\numFalse} to {\subtNumFalse}.
This increase in the false alarm rate is not surprising, since we are
looking for the weaker signals that remain after the brightest ones
have been removed.
\begin{table}[tbp]
  \label{t:subtraction}
  \begin{tabular}{||c||c|c|c|c|c|c|c|c||}
\hline
\hline
Rounds of Subtraction & 0 & 1 & 2 & 3 & 4 & 5 & 6 & 7 \\
\hline
\hline
Found Signals & 1989 & 2962 & 3250 & 3346 & 3392 & 3405 & 3417 & 3419 \\
\hline
False Alarms & 5 & 23 & 24 & 27 & 28 & 28 & 29 & 29 \\
\hline
\hline
\end{tabular}

  \caption{
    Results of iterative signal subtraction, described in \sref{s:subtract}.
  }
\end{table}

\section{Comparison to Entries in the Second Mock LISA Data Challenge}

An earlier version of this pipeline, using the long-wavelength
approximation, was used to generate our entry in the second Mock LISA
Data Challenge~\cite{MLDC2Poster}.  Several other MLDC2 entries
analyzed the same data set, as described in \cite{Babak:2007zd}.  The
parameters returned by those ``blind'' searches are recorded at
\cite{mldc2results}, and can be compared to the present pipeline using
the criteria described in \sref{s:eval}.  A script to do this is in
the \texttt{MLDCevaluation/Galaxy\_Evaluation/AEI} directory of
lisatools~\cite{lisatools}, and we have run this on the entries, along
with suitably converted versions of the searches reported in
\sref{s:results} and \sref{s:subtract}.  For brevity we only report
the numbers of false alarms and false dismissals, which are summarized
in \tref{t:evaluation}.

\begin{table}[tbp]
  \label{t:evaluation}
  \begin{tabular}{||c||c|c|c|c||c|c|c||}
\hline
\hline
Search & MTJPL & PWAEI & IMPAN & UTB & LW & RA & RAsubt \\
\hline
\hline
Found Signals & 18084 & 1766 & 264 & 281 & 1704 & 1989 & 3419 \\
\hline
False Alarms & 1240 & 11 & 140 & 3581 & 9 & 5 & 29 \\
\hline
\hline
\end{tabular}

  \caption{
    Comparison of different entries in the Second Mock LISA Data Challenge,
    and results of the current pipeline.  The MLDC2 entries, which are
    described in~\cite{Babak:2007zd} and available from 
    \cite{mldc2results}, are: 
    ``MTJPL'', a submission by 
    Crowder et al~\cite{Crowder:2006eu,Crowder:2007ft} using the
    Metropolis-Hastings Monte Carlo with a multi-signal template;
    ``PWAEI'', a submission by two of the present authors (Prix and
    Whelan) using a version of the present pipeline with the long-wavelength
    response~\cite{MLDC2Poster}; ``IMPAN'', a submission by Kr\'{o}lak and
    B{\l}aut using an $\mc{F}$-statistic method~\cite{Krolak:2004xp}
    which was refined for subsequent MLDC
    rounds~\cite{Blaut:2009zz,Blaut:2009si}; and ``UTB'', a submission by
    Nayak et al using a tomographic method to map the overall distribution
    of galactic binaries~\cite{Nayak:2007zz}.  Included for comparison are
    ``LW'', the present search with the long-wavelength response; ``RA'',
    the present search with the rigid adiabatic response; and ``RAsubt'',
    the search described in \sref{s:subtract} with seven rounds of signal
    subtraction.
  }
\end{table}

Our original MLDC entry returned the second highest number of true
signals.  While the UTB entry included more candidate signals than
ours, very few of them were associated with actual sources according
to the method of \sref{ss:eval-id}.  Incidentally, the same
qualitative information carried in the right-hand panel Figure~1 of
\cite{Babak:2007zd} is reflected in \tref{t:evaluation}, which
considers the match between Doppler parameters of candidate signals
and sources in the key: the PWAEI candidates have the lowest false
alarm percentage, most of the {\numFoundMTJPL} MTJPL signals are real,
and the UTB signals, in addition to having no amplitude parameter
information and not determining the sign of the ecliptic latitude,
have no discernible one-to-one correlation with the true sources.

The results of our original MLDC entry (``PWAEI'') are similar to the
LW results reported in \sref{s:results}, but not identical because the
original search used the LISA simulator data, and the present search
uses data from synthetic LISA.  The addition of the rigid adiabatic
response makes a drastic improvement in amplitude parameter recovery
as detailed in \sref{ss:results-ampErrors}, but also increases the
number of recovered signals, as does the experimental iterative signal
subtraction technique detailed in \sref{s:subtract}.  The various
incarnations of our pipeline recovered between {\numFoundLWL} and
{\subtNumFound} of the {\numBright} ``bright'' signals present in the data set
(which contained a further 30~million background signals assumed to be
undetectable); note, however, that due to our current pipeline
first-stage threshold of $2\F>20$ (see Sec.~\ref{s:pipeline-search})
we only expect to be able to find at most {\numForty} or so signals
with $\abs{\Akey}^2\ge 40$.
The MTJPL search by Crowder et al was able to recover {\numFoundMTJPL} true
signals. This is still less than {\numBright}, so the a priori
assessment of ``bright'' signals should be taken with a grain of salt.
However, the MTJPL pipeline also only missed {\numMissedFortyMTJPL} of
those {\numForty} signals with $\abs{\Akey}^2\ge 40$, which indicates
a higher intrinsic efficiency.

One advantage of our pipeline is speed.  The full search can run in a
matter of hours using a few hundred nodes of a computing cluster.  The
iterative signal subtraction was run over the course of a week, but
much of that was taken up in subtracting signals and re-starting the
pipeline by hand. The speed will also be affected by the choice of
first-stage threshold in our pipeline (currently $2\F>20$). We could
increase the number of signals found by lowering this threshold, but
the next-stage follow-up steps and signal-subtraction would then take
more time.
More work would be required to understand how much efficiency could be
gained at what computing cost by lowering this threshold, and how it
would affect the quality and reliablity of signal extraction.

\section{Conclusions}

We have applied an $\F$-statistic template bank search to mock LISA data
containing a full galaxy of simulated white-dwarf binary systems.  A
multi-stage pipeline requiring Doppler parameter {\coinc} between
searches using different TDI variables is effective in distinguishing
true signals from false alarms and allows {\numFound} signals to be
recovered with only {\numFalse} false alarms.

The use of the rigid adiabatic model for LISA response, including a
response tensor depending on signal frequency and sky direction,
eliminates the systematic amplitude parameter errors associated with
searches using a long-wavelength approximation, and also allows more
signals to be identified than with the simpler long-wavelength
response tensor.

The relatively accurate recovery of both amplitude and Doppler
parameters allows an experimental implementation of an iterative
signal subtraction pipeline; after {\subtRounds} rounds of signal
subtraction and re-analysis, the number of found signals was increased
to {\subtNumFound}, with a total of {\subtNumFalse} false alarms.

\acknowledgments

We thank Stas Babak, Curt Cutler, Ilya Mandel, Michele Vallisneri, and
Alberto Vecchio and for helpful discussions and comments.
This work was supported by the Max-Planck-Society, by DFG grant
SFB/TR~7, by the German Aerospace Center (DLR), by NSF grant PHY-0855494,
and by the College of
Science of Rochester Institute of Technology.  DK would like to thank
the Max Planck Institute for Gravitational Physics (Albert Einstein
Institute) for support and hospitality.  JTW also wishes to thank AEI
Potsdam, which was his home institution during much of this project.
The analysis for this project was performed on the Morgane cluster at
AEI Potsdam and the Atlas cluster at AEI Hannover.
This paper has been assigned LIGO Document Number \dcc.

\bibliography{biblio}

\end{document}